\documentclass[aps,prl,twocolumn,groupedaddress,showpacs]{revtex4}
\usepackage{graphicx}
\usepackage{float}
\usepackage{dcolumn}
\usepackage{amsmath}
\usepackage{amssymb}
\input{epsf}

\newtheorem{theorem}{Theorem}

\newtheorem{claim}[theorem]{Claim}
\newtheorem{lemma}{Lemma}

\newtheorem{corollary}[theorem]{Corollary}

\newcommand{\qedsymb}{\hfill{\rule{2mm}{2mm}}}
\newenvironment{proof}[1][]{\begin{trivlist}
\item[\hspace{\labelsep}{\bf\noindent Proof#1:\/}] }{\qedsymb\end{trivlist}}

\newcommand{\ket}[1]{\left| {#1} \right\rangle}
\newcommand{\bra}[1]{\left\langle {#1}\right |}
\newcommand{\iprod}[2]{\left\langle {#1}|{#2}\right\rangle}

\def\Tr{{\rm Tr}}

\def\be{\begin{equation}}
\def\ee{\end{equation}}
\def\bea{\begin{eqnarray}}
\def\eea{\end{eqnarray}}

\def\>{\rangle}
\def\<{\langle}

\begin{document}
\title{
\[ \vspace{-2cm} \]
\noindent\hfill\hbox to 1.5in{\rm  } \vskip 1pt \noindent\hfill\hbox
to 1.5in{\rm SLAC-PUB-12039 \hfill  } \vskip 1pt
\noindent\hfill\hbox to 1.5in{\rm August 1, 2006 \hfill}\vskip 10pt
Exploring Contractor Renormalization: Tests on the 2-D Heisenberg
Antiferromagnet and Some New Perspectives \footnote{This work was
supported by the U.~S.~DOE, Contract No.~DE-AC02-76SF00515.}}
\author{M. Stewart Siu\footnote{msiu@stanford.edu} and Marvin Weinstein\footnote{niv@slac.stanford.edu}}
\address{Stanford Linear Accelerator Center, Stanford University,
  Stanford, California 94309}

\begin{abstract}

Contractor Renormalization (CORE) is a numerical renormalization
method for Hamiltonian systems that has found applications in
particle and condensed matter physics. There have been few studies,
however, on further understanding of what exactly it does and its
convergence properties. The current work has two main objectives.
First, we wish to investigate the convergence of the cluster
expansion for a two-dimensional Heisenberg Antiferromagnet(HAF).
This is important because the linked cluster expansion used to
evaluate this formula non-perturbatively is not controlled by a
small parameter. Here we present a study of three different blocking
schemes which reveals some surprises and in particular, leads us to
suggest a scheme for defining successive terms in the cluster
expansion. Our second goal is to present some new perspectives on
CORE in light of recent developments to make it accessible to more
researchers, including those in Quantum Information Science. We make
some comparison to entanglement-based approaches and discuss how it
may be possible to improve or generalize the method.

\end{abstract}

\maketitle


\section{1. Introduction and Outline}

Many problems in physics, in areas ranging from particle and
condensed matter physics to theoretical quantum computing, can only
be treated by numerical methods.  Among them is the particularly
interesting problem of extracting the low energy behavior of a
multi-dimensional system defined by a Hamiltonian with local
interactions. While analytical methods can be applied to a few such
Hamiltonians, existing methods generally require enormous
computational power to study systems of even modest size. For
example, Quantum Monte Carlo can give highly accurate results for
many systems, but its applicability can be limited by the fermion
sign problem, as well as the inaccuracies inherent in extrapolating
finite size results to the limit of infinite volume. The most
popular alternative, the Density Matrix Renormalization Group method
(DMRG)\cite{densitymatrix}, is particularly successful in one
dimension.  With the advent of Quantum Information Science, it has
been extended to simulate time evolution \cite{TEBD} and
multidimensional models (PEPS or tensor networks\cite{PEPS, TN}).
The idea underlying DMRG and its generalizations is a clever
variational ansatz - the representation of states as contracted
tensors. This approach has many virtues. For example, it provides an
upper bound on the ground state energy density. One can also extract
quantities such as total entropy which is approximately preserved in
the subspace and use them to analyze phase transition
\cite{LegezaPT, Latorre}. It has its limitations, however, as one
has to deal with convergence issues and a very large number of
states in more complicated systems. The closely-related method
described in \cite{PEPS} also has the requirement of a fixed finite
lattice.

In search for an alternative method, one might reasonably ask if
there is a way that encodes the information about the ground state
and low lying excited states not in states, but in \em operators
\em. One method that does that is the Contractor Renormalization
Group Method (CORE)\cite{COREpaper}, which, like DMRG, is an attempt
to improve Wilson's real-space renormalization\cite{Wilson}. Similar
ideas have appeared in the past \cite{Cloizeaux, Zivkovic} but we
will follow the terminology of \cite{COREpaper} as it is the first
general formulation of the method. (The same formulation was also
independently proposed as Real-Space Renormalization Group with
Effective Interaction (RSRG-EI) in \cite{MalrieuGuihery}). While the
method has intuitive appeal and has been used to study many
collective phenomena\cite{COREpaper2,calzado,somepapers}, unlike in
the case of DMRG, we have relatively little grasp of exactly why and
when the algorithm works. There seems to be a need to put aside
applications for a moment and look at the method more closely. Short
of very conclusive results, we present here some findings that may
lead to a better understanding of CORE. We approach the issue from
two angles. First, we take the simplest 2-D model as a testing
ground and carry out successive numerical renormalization using
three different blocking schemes. Our choice, the Heisenberg
Antiferromagnet (HAF), has been studied with RSRG-EI in the past
\cite{MalrieuGuihery}, but our focus is on the extraction of order
parameter and the behavior of long-range operators in the 2-D
cluster expansion, particularly because there is a freedom in
defining successive terms in the expansion \cite{LegezaOrder}. We
also look at how well different blocking schemes agree with each
other while capturing the physics in distinct ways. Secondly, we
make some theoretical and numerical comparison between CORE and DMRG
as well as Entanglement Renormalization \cite{DU}. By presenting a
slightly different perspective, we hope to provoke further
investigation into the limitations of the method and the possibility
for improvement.

Section 2 of the paper reviews the basic formulas in CORE and sets
the notation. The first subsection of Section~3 presents calculation
of CORE on nine-site blocks, as considered in \cite{MalrieuGuihery}.
In addition to the energy density, which can be easily obtained to
high accuracy, we calculate the staggered magnetization and find
that, without longer-range operators, it is only accurate to one
significant figure. To see how longer-range terms behave with
limited computing resource, we use smaller blocks in section 3(b)
(four-site) and 3(c) (five-site) and compute the energy density. We
find that operators beyond nearest-neighbor can contribute quite
significantly and requires very careful ordering. To this end we
propose a ordering scheme based on diameters of the clusters. The
effect of long-range operator, however, does not necessarily
correspond to long-range correlation, as we see in 3(b) that even
with only nearest neighbor operators a vanishing mass gap appears
non-trivially. Although application is not the focus of the current
work, we also show as a side note in Section 3(c) how, under the
appropriate blocking, CORE might provide an interesting
justification for the spin-wave approximation.

In Section 4 we turn to the question of how CORE relates to
entanglement-based approaches. In the first subsection we discuss
truncation and blocking schemes of CORE and DMRG and compare their
principles. We show that for a small, finite toy model CORE yields
results comparable to DMRG. Then in Section 4(b) we formulate CORE
in a way that enables us to see its similarity to Entanglement
Renormalization\cite{DU}, which should be more familiar to readers
in Quantum Information Science. We also discuss in Section 4(c) the
role of block entropy in CORE and its possible use. Finally in
Section 4(d) we discuss how the choice of retained states in CORE
can affect its performance and what entanglement has to do with this
choice.

In Section 5 we reprise our results and discuss a number of future
directions and possibilities.

\section{2. CORE: The Basic Formulas}

The original description of CORE can be found in
Ref.\cite{COREpaper}. This section summarizes the basic formulas we
will use in the sections to follow (particularly in Section 4).

Given a Hamiltonian $H$ on Hilbert space $\cal H$, the renormalized
Hamiltonian that CORE seeks to approximate takes the form:
\begin{eqnarray}
H_{ren}(t) &=& \nonumber \\
&&\hskip-35pt
(Pe^{-2Ht}P)^{-\frac{1}{2}}Pe^{-Ht}He^{-Ht}P(Pe^{-2Ht}P)^{-\frac{1}{2}}
.
\nonumber\\
\label{Hrent}
\end{eqnarray}
where $P$ is a projection operator from $\cal H$ onto $\cal H'$, a
chosen subspace of retained states (in the language of
RSRG-EI\cite{MalrieuGuihery}, the model space). $t$ is a variable
parameter usually taken to infinity. With the following lemma, we
can show that $H_{ren}=\lim_{t\to\infty} H_{ren}(t)$ takes a
particularly simple form.

\begin{lemma} Let $\ket{a_i}, i=1,\ldots,M$ be an arbitrary orthonormal
basis for ${\cal H'}$ and let $\ket{v_l}, l=1,\ldots,N$, be an
orthonormal set of eigenvectors of the Hamiltonian, $H$, with
eigenvalues $\{E_1,E_2,\ldots, E_l \ldots\}$ arranged in ascending
order (generally, $M < N$). Then, there exists an $M \times M$
matrix, $R_{ij}$, such that the states $\ket{w_j} = \sum_{i=1}^{M}
R_{ij} \ket{a_i} $ have the property that for each $j$, there is
exactly one index $1\leq f(j)\leq N$ such that
$\iprod{w_j}{v_{f(j)}}\neq 0$ and $\iprod{w_k}{v_{f(j)}}=0\, \forall
k>j$. \label{lemmaofR}
\end{lemma}
\begin{corollary}
Each state $\ket{w_j}$ has the property that \be
    e^{-Ht}\ket{w_j}\sim_{t\to\infty}
    e^{-E_{f(j)}t}\iprod{v_{f(j)}}{w_j}\ket{v_{f(j)}} + \ldots ,
\label{contract} \ee where $\ldots$ stands for terms that vanish
more rapidly as $t \to \infty$.
\end{corollary}
Then it can be proved that:

\begin{eqnarray}
H_{ren} &=& \sum_{i,j,k} \ket{a_j} R^{\dag}_{ji} E_{f(i)} R_{ik}
\bra{a_k} ,\nonumber \\
 &=& \sum_{i} \ket{w_i} E_{f(i)} \bra{w_i}
\label{diagform}.
\end{eqnarray}

On a lattice we may expand this renormalized Hamiltonian as linked
clusters:
\begin{eqnarray}
H_{ren} = \sum_{s\subset L} h^{conn}(s), \nonumber \\
h^{conn}(s)\equiv H_{ren}\mid_s - \sum_{s'\subset s}h^{conn}(s')
\label{fce}
\end{eqnarray}
where $L$ is the entire lattice corresponding to $\cal H'$ (it can
be infinite) and $s$ is a connected sublattice. $H_{ren}\mid_s$ is
$H_{ren}$ evaluated for the theory obtained by restricting the full
Hamiltonian to the sublattice $s$. For notational convenience,
operators acting on $s$ are implicitly assumed to extend to the full
lattice by acting as the identity operator on degrees of freedom
lying outside of $s$. We will also refer to $h^{conn}(s)$ as a
range-$r$ connected operator if $s$ contains $r$ blocks used to
define the projection $P$, and we will adopt the notation
$h^{conn}(s^r)$ to denote such an operator.

In practice we can of course only evaluate range-$r$ operators for
small $r$, but since the longer-range terms are not controlled by a
small parameter (this would be true if $t$ is a small number), we
would have to be careful what to discard and we will discuss some
examples in the next section.

\section{3. The Two-Dimensional Heisenberg Antiferromagnet}

We have chosen the 2-D spin-1/2 HAF as testing ground for its
simplicity and connection to the Hubbard model. The Hamiltonian is
defined as\bea
H&=&\sum_{<i,j>} \vec{S}_i\cdot \vec{S}_j \nonumber \\
&=&\sum_{<i,j>} S^x_i \cdot S^x_j + S^y_i \cdot S^y_j + S^z_i \cdot
S^z_j , \eea where the sum is over all neighboring pairs on a
two-dimensional square lattice.

\subsection{a. Nine-site Square Blocking}

Although we are not able to compute (on a PC) long-range terms with
nine-site blocks, their simplicity allows us to see some basic
features of the computation and the HAF we are studying. The
nine-site block is a natural extension of the three-site block in
1-D \cite{COREpaper} as it also has a spin-1/2 multiplet as ground
states which we use as retained states. The range-1 and range-2
operators take the form: \bea
h^{conn} (s^1)=\alpha_0 \mathbf{1} \nonumber \\
h^{conn} (s^2)=\alpha_1 \mathbf{1} + \beta_1 \vec{S} \cdot \vec{S} .
\label{rg9s} \eea where we denote the one-block and
two-adjacent-block configurations by $s^1$ and $s^2$ respectively.
Note that no calculation of $R$ in Eq.\ref{lemmaofR} is necessary to
find $h^{conn}(s^2)$, because the retained states form exactly a
spin-0 multiplet and spin-1 multiplet. If we are to preserve the
spin symmetry, there are no other inequivalent rotations. The fact
that spin symmetry largely dictates possible eigenvectors of the
renormalized Hamiltonian greatly reduces computational effort,
though arguably, we can only test the effect of CORE's recipe for
$R$ in more complicated situations where there are many multiplets
of the same spin. (From our experience it is also possible to break
the spin symmetry and let CORE decide the best linear combination.
This often yields a good ground state energy but a poor spectral
distribution.)

\begin{figure}
  \includegraphics[width=3in]{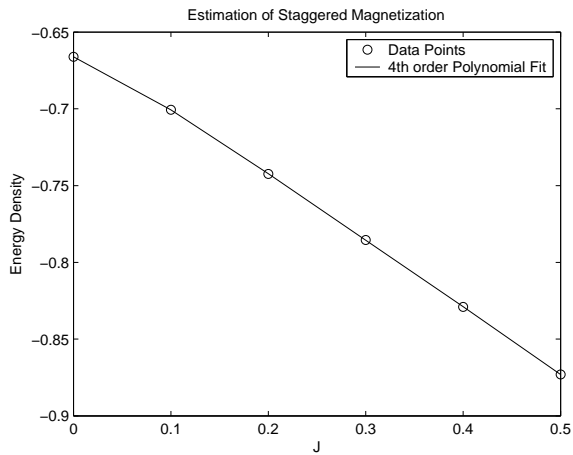}\\
  \caption{A plot of staggered magnetic field J vs energy density E.
  Staggered magnetization can be calculated from $\langle M\rangle=\frac{\delta
\langle H+JM \rangle}{\delta J} \mid_{J=0}$. The data points have to
be close enough to see the curvature but sparse enough so that error
in the energy will not affect the slope between the points too
significantly.} \label{mag}
\end{figure}

With the range-1 and range-2 operators above, we can easily
calculate the ground state energy density by summing a geometric
series and reproduce the result of \cite{Zivkovic, MalrieuGuihery}.
The energy per site obtained this way is $-0.666$ and within $0.5\%$
of the Monte Carlo result $-0.669$\cite{exact}. The situation
becomes much more complicated, however, when we we proceed to
calculate the staggered magnetization, an order parameter of the
HAF. There are two ways of calculating the expectation value of a
renormalized operator. In \cite{COREpaper} it is argued that other
operators should take a form similar to Eq.\ref{diagform}: \be
   O_{ren}=R^\dag O R,
\ee where now $O$ is the matrix in the $\ket{w_i}$ basis \be
    O_{ij} = \bra{w_i} O \ket{w_j}.
\ee But since the staggered magnetization does not easily converge
to a simple form as the Hamiltonian does, it is easier to calculate
the expectation value by: \be {\langle O\rangle}=\frac{\delta
\langle H+JO \rangle}{\delta J} \mid_{J=0} \label{Jthm} \ee. In
other words, we add a multiple of the staggered magnetization
operator, $J M_{stag}$, to the Hamiltonian, use CORE to compute the
ground state energy density, $E(J)$, and extract the slope of this
function at $J=0$(see Eq.\ref{Jthm}). Once we have added $J
M_{stag}$ to the starting Hamiltonian the renormalized Hamiltonian
is no longer a simple multiple of the original Hamiltonian and
obtaining $E(J)$ requires running the RG until it converges. In
Fig.\ref{mag} we plot the ground state energy obtained in this way
as a function of $J$. The staggered magnetization is the slope of
the curve at $J=0$ and is obtained by fitting to a fourth-order
polynomial in $J$.

How small can we set the $J$-values to be? If we are allowed to use
the knowledge that the energy obtained is accurate to two
significant figures, we have to choose the $J$-values to be large
enough so that the error in the slope will be at least one order of
magnitude smaller than the slope itself. The expectation value thus
obtained is $M_{stag}= -0.29$, whereas the exact Monte Carlo result
is $-0.30$\cite{exact}. We must caution though that if we use
smaller $J$'s or change the order of the polynomial fit, we can
change the result by up to $20\%$. To calculate the order parameter
more reliably, it seems that one has to find a way to capture more
physical information in the renormalization.

\subsection{b. Four-site Square blocking}

In the case of the four-site block the lowest lying states form a
spin-0 singlet and a spin-1 triplet.  In this case, the renormalized
Hamiltonian is no longer isomorphic, even at range-2, to the
original Hamiltonian which had a single spin-1/2 degree of freedom
associated with each site.  Furthermore, it appears to describe a
theory with a non-vanishing mass gap, as subsequent RG-steps
continues to have a spin-0 and a spin-1 multiplet as its lowest
energy eigenstates. This single-site gap, however, is just a
reflection of the uncertainty principle cost one must pay for
localizing the spin-1 excitation to a single block. By keeping these
two multiplets at every step, we run the RG until the energy density
converges and find that the gap between the two multiplets converges
to zero. This reflects the fact that spin-1/2 HAF is massless and
agrees with the result of nine-site and five-site blockings, where
the ground state is by contruction degenerate. We would like to note
that this result is entirely non-trivial, since the original
Wilsonian RSRG (the $t=0$ limit of the CORE formula for the
renormalized Hamiltonian) predicts that the same theory dimerizes
and has a non-vanishing mass gap.

While this result gives us some confidence that even the lowest
order cluster expansion is doing fairly well at extracting the
correct physics, we obtain $-0.710$ for the ground state energy
density - it is not nearly as accurate as the nine-site case. This
is perhaps to be expected given that we have exactly diagonalized a
mere $256$ states here (whereas we have diagonalized $2^{18}$ states
in the nine-site calculation), yet in some sense it is still
remarkable, since we have kept proportionally more states in every
iteration (4-out-of-16 vs 2-out-of-512). To check the convergence of
the cluster expansion, we add the operator corresponding to a
sub-cluster consisting of four blocks arranged in a square. We do
this only for the first RG step as the analogous term in the next
iteration requires diagonalization of $2^{32}$ states. This already
improves the energy density to $-0.677$ and allows us to obtain a
staggered magnetization that is close to the nine-site case. At this
point we might be prompted to ask: To achieve a good accuracy, do we
simply compute all the terms within our computational power? Since
the five-site blocking exhibits the most numerical sensitivity to
the ordering in the cluster expansion, we defer a more detailed
discussion to the next subsection.

\begin{figure}
  \includegraphics[width=3in]{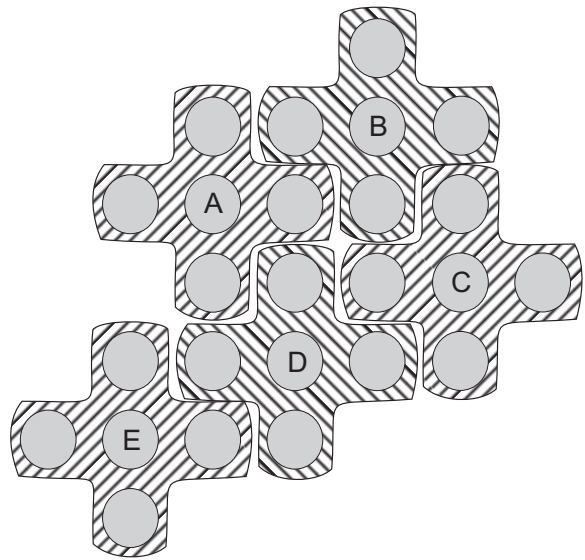}\\
  \caption{Blocking of five-site stars. A total of five blocks are shown in the figure.
  range-2 terms are connected operators on two adajcent blocks, such as A and B. There are two
  types of range-3 terms: Examples of "L"s include ABC, BCD, etc., while there is a
  straight-line term on CDE. Finally, the range-4 plaquette acts on ABCD. }\label{star}
\end{figure}

\subsection{c. Five-site Star blocking}

The final CORE computation we wish to discuss uses the five-site
blocks shown in Fig.\ref{star}. What makes the RG transformation
following from this blocking procedure interesting is that it
behaves quite differently from the one obtained by restricting
attention to square blocks. This is because the ground state of a
star, in contrast to the nine-site block, is a spin-3/2 multiplet.
The renormalized Hamiltonian at range two takes the general form
\begin{equation}
h^{conn}(s^2)=c_0 \mathbf{1} + c_1 \vec{S} \cdot \vec{S} +c_2
(\vec{S} \cdot \vec{S})^2 + c_3 (\vec{S} \cdot \vec{S})^3
\label{r2ss} .
\end{equation}
If we evaluate the expectation value of Eq.\ref{r2ss} in the
classical N\'eel state, we obtain an \em upper bound \em on the
energy density of -0.712 per site.  So despite the fact that the
fundamental block has more sites, this accuracy is worse than the
equivalent four-site blocking. Notice that the ratio of intra-block
over inter-block links is smaller with five-site blocks. We suspect
that the large perimeter of star blocks result in many
under-constrained sites near the edge of the configuration.

Nevertheless, we can continue to run the renormalization group to
see how the theory changes.  If we use the Hamiltonian defined in
Eq.\ref{r2ss} and once again calculate the spectrum of a single star
we find that now the ground state is a spin-9/2 multiplet. Keeping
the spin-9/2 multiplets as new degrees of freedom, we find that the
range-2 interaction again to be antiferromagnetic. We use these
interactions to construct a five-site star, and the ground state now
becomes a spin-$27/2$ multiplet. From this version of CORE we see
that the spin-$1/2$ theory is equivalent to a theory with a larger
spin at each site.

Clearly the stability of a picture that says that spins keep growing
with each renormalization group step must be checked by computing
the contributions to the renormalized Hamiltonian coming from larger
clusters.  The reason this is interesting is that once one goes to
larger sub-clusters, one introduces new diagonal couplings,
couplings involving three sites at a time, etc. Since the diagonal
couplings are also mainly antiferromagnetic in character, they tend
to introduce frustration into the system. It is entirely possible
that these are relevant operators and significantly modify the
nature of the RG flow, so that at the next step the spins fail to
grow.

To obtain a qualitative understanding of how each connected operator
modifies the flow of the renormalized Hamiltonian we first observe
that each three-site connected operator can be written as a sum of
terms which act non-trivially on one, two or three-sites at a time.
To see why this is true let us assume that we have spin-$n/2$ on
each site of the new lattice and let $X_i, i=2..n+1$ be a set of $n$
traceless matrices which, together with the matrix $X_1 = 1/(n+1)
\,{\bf 1}$ (where ${\bf 1}$ is the unit matrix), form a basis for
the space of of $(n+1) \times (n+1)$ matrices. Furthermore, assume
that these matrices satisfy the normalization conditions \be
    \Tr(X_i\,X_j) = \delta_{ij}.
\ee
It then follows that the tensor products
\be
    M_{ijk} = X_i \otimes X_j \otimes X_j
\ee
form a normalized basis of matrices which operate on three lattice sites
at a time.  Thus, the general three-site connected operator
can be written as
\be
    O = \sum_{i,j,k} \alpha_{ijk} M_{ijk}.
\ee
Furthermore, the coefficents $\alpha_{ijk}$ can be extracted
by taking the trace with any $M_{ijk}$; i.e.,
\be
    \alpha_{ijk} = \Tr(M_{ijk} O ). \ee Given this definition, it is
easy to define what we mean by the parts of $O$ which act
non-trivially on zero, one, two or all three sites.  The part of $O$
that is proportional to a multiple of the unit matrix is, of course,
a term that acts trivially on all three sites and won't contribute
to dynamics. Of what remains, it typically turns out that the
operators which act non-trivially on only one or two sites are the
most important operators.  Thus, for the purpose of getting a simple
qualitative understanding of the stability of our problem we
restrict attention to these operators in what follows. Since the
renormalized Hamiltonian must commute with the total spin operator,
it follows from direct computation, or simple group theory, that
there are no terms in the renormalized Hamiltonian which act
non-trivially on a single block.  (This is because for spin-$n/2$,
the space of $(2n + 1) \times (2n+1)$ matrices decomposes, under
total spin, into matrices which transform as spin-$0$, the unit
matrix, and, for each $j$, a set of matrices which transform as
spin-$j$, where $j$ runs from $1\ldots 2n+1$.)  Similar arguments
tell us that operators that act non-trivially on only two-sites at a
time can be written as polynomials in $\vec{S} \cdot \vec{S}$ acting
on the two-sites in question.  By $\vec{S}$ we mean the matrices
which represent the generators of spin rotations for spin-$n/2$.
Note, the order of the polynomial for the case of spin-$n/2$ is at
most $2n$.

Having said this we can ask what the effects of the terms
proportional $\vec{S} \cdot \vec{S}$ (which act on only two-sites at
a time) coming from the straight line, the L and the plaquette are
(this terminology is best explained by looking at Fig.\ref{star}.),
since these two-site operators turn out to be a significant part of
the contribution to the renormalized Hamiltonian.  There are three
important observations that we must make about these terms.

First, the antiferromagnetic interaction between diagonal sites
(e.g., BD in Fig.\ref{star}) are of the same order of magnitude as
the interaction between adjacent sites.  This means that the
contribution of larger clusters to the renormalization group
transformation generically produce a {\it frustrated\/} system.
Second, unlike the case of one dimension, where the importance of
operators coming from larger clusters falls off with the number of
blocks in a cluster, terms coming from the four-block plaquette
appear to be as important as those coming from clusters consisting
of only three-blocks. (Among the three types of terms we evaluated,
the three-block straight-line has the smallest two-site
contributions.) Third, the two-site terms coming from different
clusters often cancel each other, indicating that our result is
sensitive to how we define the cluster expansion.  To give the reader
a feeling for these three points we include the following table which
shows how the energy density changes with the inclusion of different
terms when we approximate the ground state with the Neel state at
the spin-3/2 level.

\begin{tabular}{|c|c|}
  \hline
  Terms included & Approx. energy density \\ \hline
  Only range-2 & -0.712 \\ \hline
  +"L" & -0.623 \\ \hline
  +"L"+"line" & -0.586 \\ \hline
  +"L"+plaquette & -0.637 \\ \hline
  +"L"+"line"+plaquette & -0.600 \\
  \hline
\end{tabular}

At this point it is natural to ask, ``What is the best way to
arrange the terms in the cluster expansion?''. Intuitively, we
expect the contribution of a long chain in a straight line to be
less than the contribution of a square with the same number of
blocks, because the blocks in the latter case are closer to each
other. Taking this with the observations above, we propose a cluster
re-summation that treats all clusters having the same \em diameter
\em as having equal weight. By the diameter of a cluster we mean the
longest distance between two blocks in the configuration. For
example, the diameter for the range-2 term would be one, for the
three-block "L" and four-block plaquette term $\sqrt{2}$, etc. This
scheme has the additional advantage that now each term in the
cluster expansion preserves at least some of the rotational symmetry
of the lattice since, at every order, we include all the terms with
the same diameter.

Note that in a diameter expansion the four-block plaquette comes
before the three-block straight-line, which has diameter two. Of
course, to be sure that all terms coming from clusters of diameter
$2$ are smaller than those of diameter $1$ and diameter $\sqrt{2}$,
we should have also evaluated the cluster where four blocks are
arranged in a "T". We left it out because, for symmetry reasons,
this configuration is considerably more difficult to evaluate than
the plaquette and we wanted to limit this study to computations
easily carried out on an average personal computer. Using only the
plaquette and "L" terms, the ground state of the five-site spin-3/2
star again turns out to be a spin-9/2 multiplet. This is somewhat
remarkable. Even with pure $\vec{S} \cdot \vec{S}$ type interactions
between adjacent and diagonal sites, the ground state of the
generically frustrated five-site star could have been spin-1/2,
spin-3/2 or spin-9/2 depending on relative strengths of the
interactions.

In principle we should now proceed to calculate the interactions for
the spin-9/2 theory up to the same diameter. Unfortunately,
calculations of the plaquette in the spin-9/2 theory requires the
diagonalization of a sparse $4^{20}\times 4^{20}$ matrix, so we are
restricted to range-2 (diameter-1) as in the four-site case. The
range-2 interaction remains antiferromagnetic but by itself gives an
energy density that is less than $-0.78$.

Although, due to limited resources, we do not have the longer range
terms to show a converging result, we would like to close this
discussion by noting that this is very interesting in the context of
the spin-wave approximation to the HAF. Since turning the spins into
Holstein-Primakoff bosons rely on an expansion in $1/2S$, a
validated picture of growing spin essentially explains the
well-known accuracy of the spin-wave method on the spin-1/2 lattice.
We can estimate the error in the $1/2S$ expansion for a finite
spin-9/2 lattice. We take our range-2-only calculation and rewrite
the Hamiltonian on one five-site block of spin-9/2 using
Holstein-Primakoff bosons, keeping only oscillator terms that are of
no higher power than the number operator. We then find the ground
state energy by minimizing over Bogoliubov transformation
parameters. The value calculated this way differs only by $0.02\%$
from the exact energy of the spin-27/2 multiplet.

In conclusion, the three-blocking schemes give us three different
physical pictures for the same system.  The sensitivity of CORE to
link structures suggests that, when working with exotic blocking
schemes, one has to check very carefully the convergence of the
cluster expansion. Moreover, our experimentation with various ways
of resumming range-3 and range-4 terms indicates that the RG-flow
can be changed dramatically depending on how we group and order the
operators in the cluster expansion. The \em diameter \em expansion
we proposed appears to be the most plausible solution, but in
absence of rigorous theorems bounding the long-range terms, this
remains an open problem.

\section{4. Comparison to Entanglement-based Approaches}

\subsection{a. Density Matrix Renormalization}

\begin{figure}
  \includegraphics[width=2in]{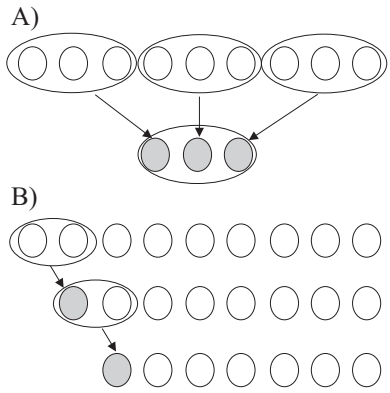}\\
  \caption{A) Hierarchical Blocking allows renormalization in the
  Wilsonian sense. B) Linear Blocking instead grows the block by one site at a time and
   works better for the original DMRG. (Details about the environment omitted above.)}
   \label{blocking}
\end{figure}

The remainder of this paper is devoted to comparing CORE to other
methods in a way we hope would shed light on the features of CORE.
Given the popularity of DMRG, we take it as an instructive benchmark
for studying CORE's performance. This sort of comparison has not
been made in the past because the two methods in their respective
original formulations have very different blocking and truncation
schemes and it seemed difficult to compare them in a meaningful way.
There are two essential features in the original formulation of DMRG
(we refer the readers to Refs.\cite{densitymatrix} for details):
\begin{itemize}
\item[\textbf{I}.] Reduced Density Matrix Truncation -
In each block we select what to keep according to the reduced
density matrix of a target state, usually the ground state of a
larger system. The larger system is the block whose truncation we
are interested in plus an "environment", which can be for example a
copy of the block. Truncation consists of eliminating those vectors
with the smallest eigenvalues of the reduced density matrix obtained
by tracing out the environment.  The error in the truncation depends
on the distribution of eigenvalues of the reduced density matrix and
can be analyzed in the language of quantum data
compression\cite{LegezaEnt}.
\\
\item[\textbf{II}.] Linear Blocking -
The block size increases linearly one site at a time. This naturally
allows iterative improvement by back and forth sweeps and gives rise
to the underlying matrix product state structure. Yet this is not
renormalization in the sense of coarse-graining one theory to
another of a different scale. To do so one could, of course, use
hierarchical blocking (Fig.\ref{blocking}A) along with the reduced
density matrix truncation described in \textbf{I} (We will refer to
this as "Hierarchical DMRG" or HDMRG), but it is usually not
preferred. Roughly speaking, the reason is that entanglement often
scales with the boundary of the block, and because hierarchical
blocking gives rise to more boundaries, truncation in HDMRG results
in more loss of information.
\end{itemize}

CORE as formulated in Section 2 does not rely on specific truncation
and blocking schemes - they are details of the projection operator
$P$. The scheme we use in Section 3 can be classified as
"hierarchical blocking" (Fig.\ref{blocking}A) and we will refer to
it as such. (Though unlike HDMRG or naive real-space
renormalization, the state cannot be reconstructed in the original
space in a simple hierarchical manner.)

Since we are free are to choose the form of the projection operator,
this difference between CORE and DMRG can be considered a
superficial one. As we mentioned in the Introduction, the essential
feature of CORE is that it encodes some information in operators
instead of states. To see this in a formulation that allows direct
comparison, let us consider the alternative form of DMRG - the
matrix product states (MPS) formalism. One common way to obtain the
ground state energy using MPS is by applying an imaginary time
evolution to a simple starting state, which we will call
$\ket{\psi}$. Suzuki-Trotter decomposition can be used to decompose
the evolution into small steps and after every step the state will
be truncated to make sure that it lies within the subspace spanned
by MPS of a fixed dimension. If the procedure converges successfully
to the desired attractor and $P$ is a projection onto the MPS
subspace, the final state should be the same as
$Pe^{-Ht}\ket{\psi}=Pe^{-Ht}P\ket{\psi}$ where $t$ is taken to
infinity and the equality follows from the fact that the starting
state lies within the projected subspace. The ground state energy is
therefore:
\begin{eqnarray}
   E_0=\frac{\bra{\psi}Pe^{-Ht}PHPe^{-Ht}P\ket{\psi}}{\bra{\psi}Pe^{-Ht}Pe^{-Ht}P\ket{\psi}}
\end{eqnarray}
This means $E_0$ is the smallest eigenvalue of the Hamiltonian:
\begin{align}
&H'_{ren}(t)= \nonumber \\
&(Pe^{-Ht}Pe^{-Ht}P)^{-\frac{1}{2}}Pe^{-Ht}PHPe^{-Ht}P(Pe^{-Ht}Pe^{-Ht}P)^{-\frac{1}{2}}
\nonumber \\
\label{HrenMPS}
\end{align}

We can now contrast this directly with Eq.\ref{Hrent}. $H_{ren}(t)$
in Eq.\ref{Hrent} contains the exact ground state energy but cannot
be evaluated exactly, therefore we have to throw away the long-range
terms caused by the diffusion of $e^{-Ht}$. When the additional
projections are inserted in $H'_{ren}(t)$ of Eq.\ref{HrenMPS}, it no
longer contains the exact ground state energy, but its ground state
energy can be found exactly. In this case the burden of a good
approximation is shifted entirely to a clever choice of $P$.

Having seen an abstract comparison, let us now return to the
original form of DMRG and compare some numbers. Because hierarchical
blocking allows CORE to handle very large lattices and DMRG requires
linear blocking, one might expect CORE's strength to be with large
systems. We were surprised to find, however, that even on small
finite lattices, CORE achieves accuracies that are comparable to
DMRG.  We demonstrate this by running HDMRG, DMRG and CORE on a
short periodic Ising chain.

\begin{figure}
  \includegraphics[width=3.4in]{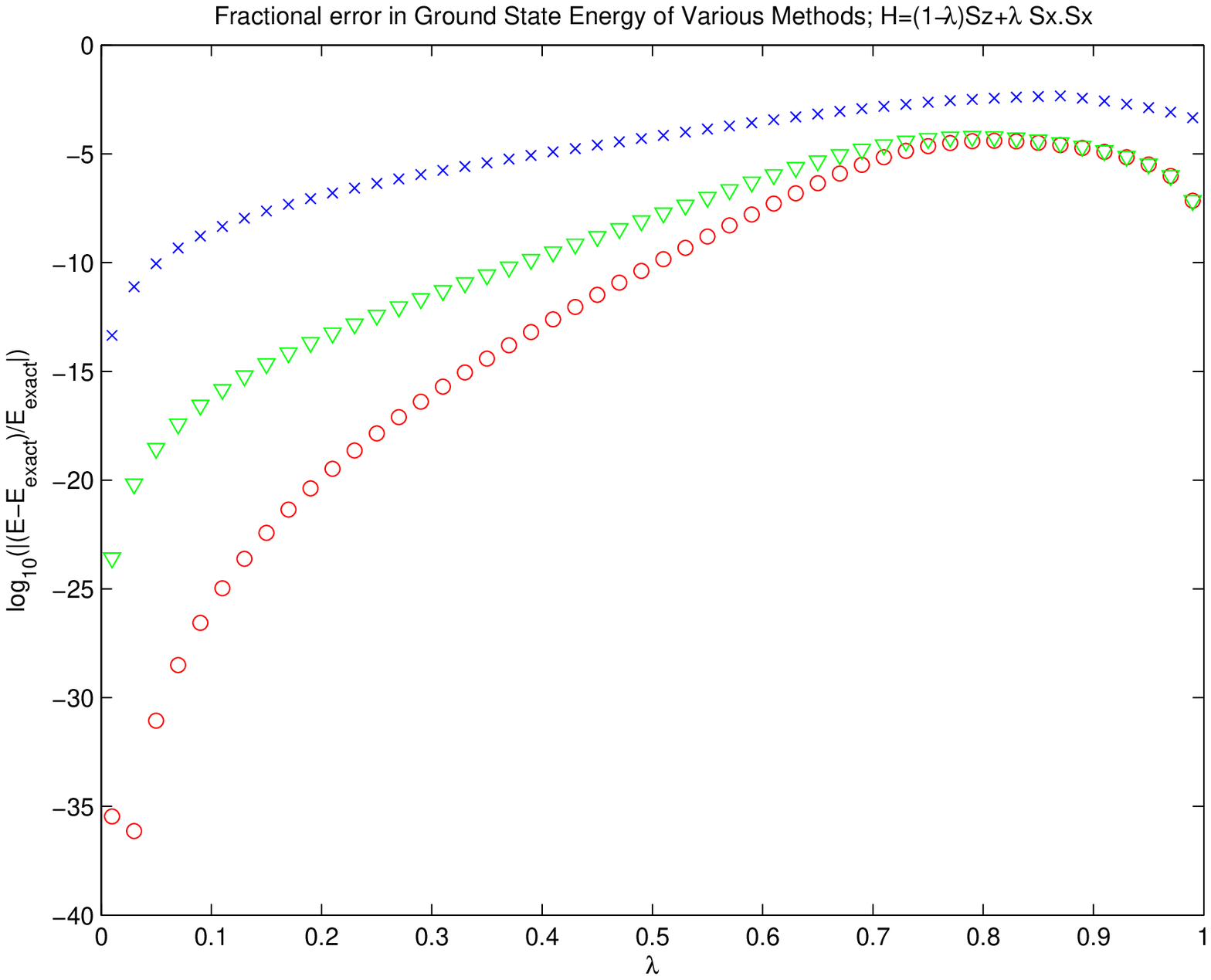}\\
  \includegraphics[width=3.4in]{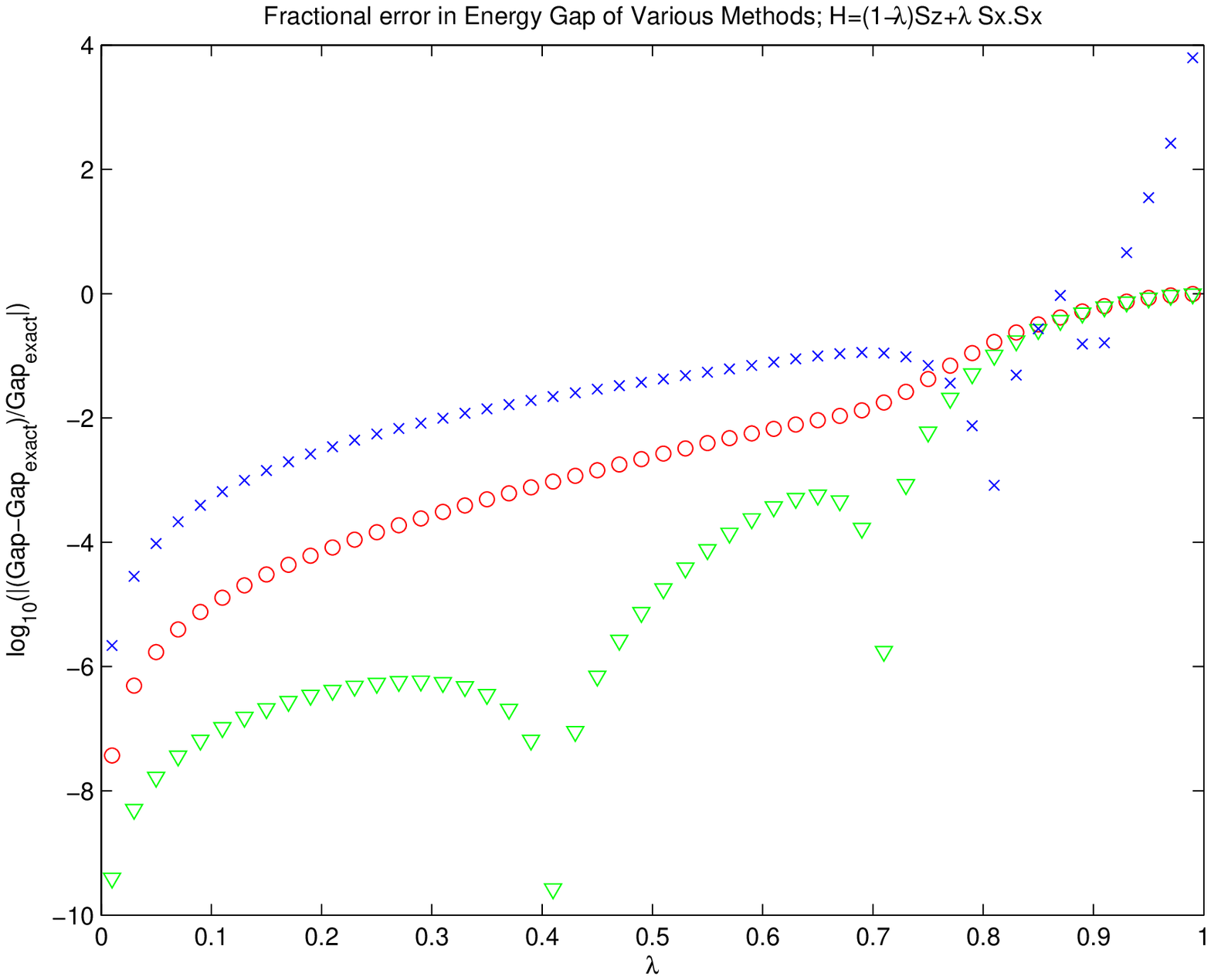}\\
  \includegraphics[width=3.2in]{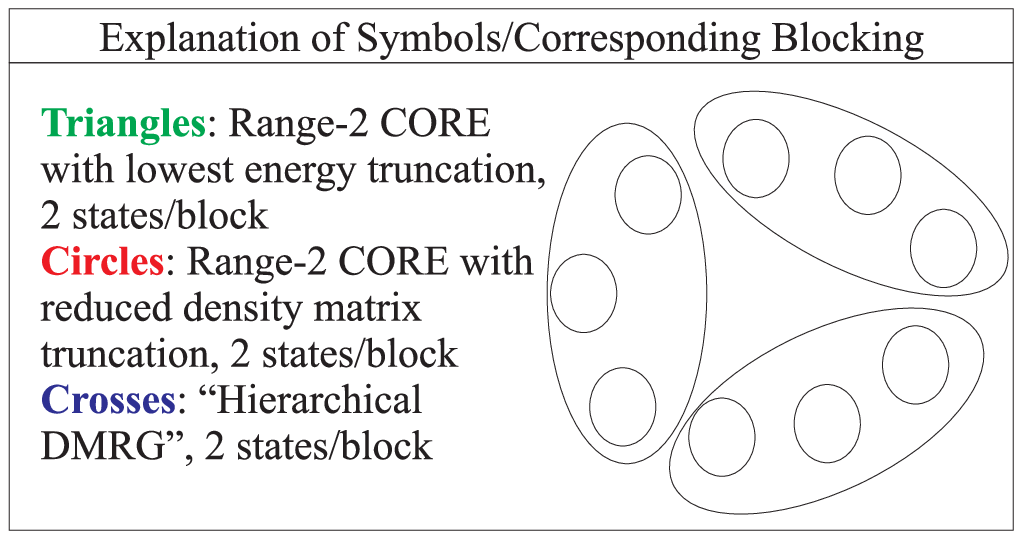}\\
  \caption{Performance of various methods on a nine-site periodic Ising
  system. The two CORE cases (triangles and circles) use different
  truncation schemes. One keeps the lowest energy states within the
  block; another finds the ground state of two blocks and considers
  its reduced density matrix. They are compared to HDMRG
  with the same blocking scheme.
  }\label{ising}
\end{figure}

The first set of plots, Fig.\ref{ising}, is for the nine-site
periodic Ising chain and compares CORE to HDMRG. This is an
interesting comparison since we can naturally use hierarchical
blocking for both. The first plot is for the ground state energy and
the second one is for the gap between the ground state and the first
excited state. The periodic chain is divided into three blocks and
we retain two states per block. To produce the HDMRG plot, we
exactly diagonalize all nine-sites to generate the target state. In
other words, we take the exact ground state, trace out six sites and
use the reduced density matrix on three sites to determine what to
keep in each block.  At the end we diagonalize the Hamiltonian
truncated to the retained states to produce the numbers for the
plot. Of course, in a realistic calculation, we do not have the
exact ground state to use as target state, but it can be
approximated by iteration, so we are essentially simulating the
ideal converged limit.

CORE calculations are carried out up to range-$2$ in the same manner
as in Section 3. (Since we are dealing with a finite lattice,
range-$3$ would be exact.) In the renormalized space we'd have three
sites and a new nearest-neighbor Hamiltonian, which we can use to
find the energy and the gap. Actually, there are two versions of
CORE in our plots with different $P$ operators. We will elaborate on
this further in Section 4(d). The first version (represented by
triangles) retains the two states with the lowest energies in each
block just as in Section 3. The second version (represented by
circles) uses a reduced density matrix truncation scheme: it takes
the ground state of six sites, trace out three sites on one side,
and retain the two states with the highest weight in the reduced
density matrix. The plot indicates that the second version performs
a little better in energy but at the expense of the gap. Note that
the six-site calculation is only one out of many ways to obtain a
reduced density matrix. In fact if we use the ground state of all
nine sites as in our HDMRG calculation instead, the states with the
highest density matrix weight turns out to be the same as those of
the first version. We have chosen the six-site case because they are
what we use for the range-$2$ operator calculation.

\begin{figure}
  \includegraphics[width=3.4in]{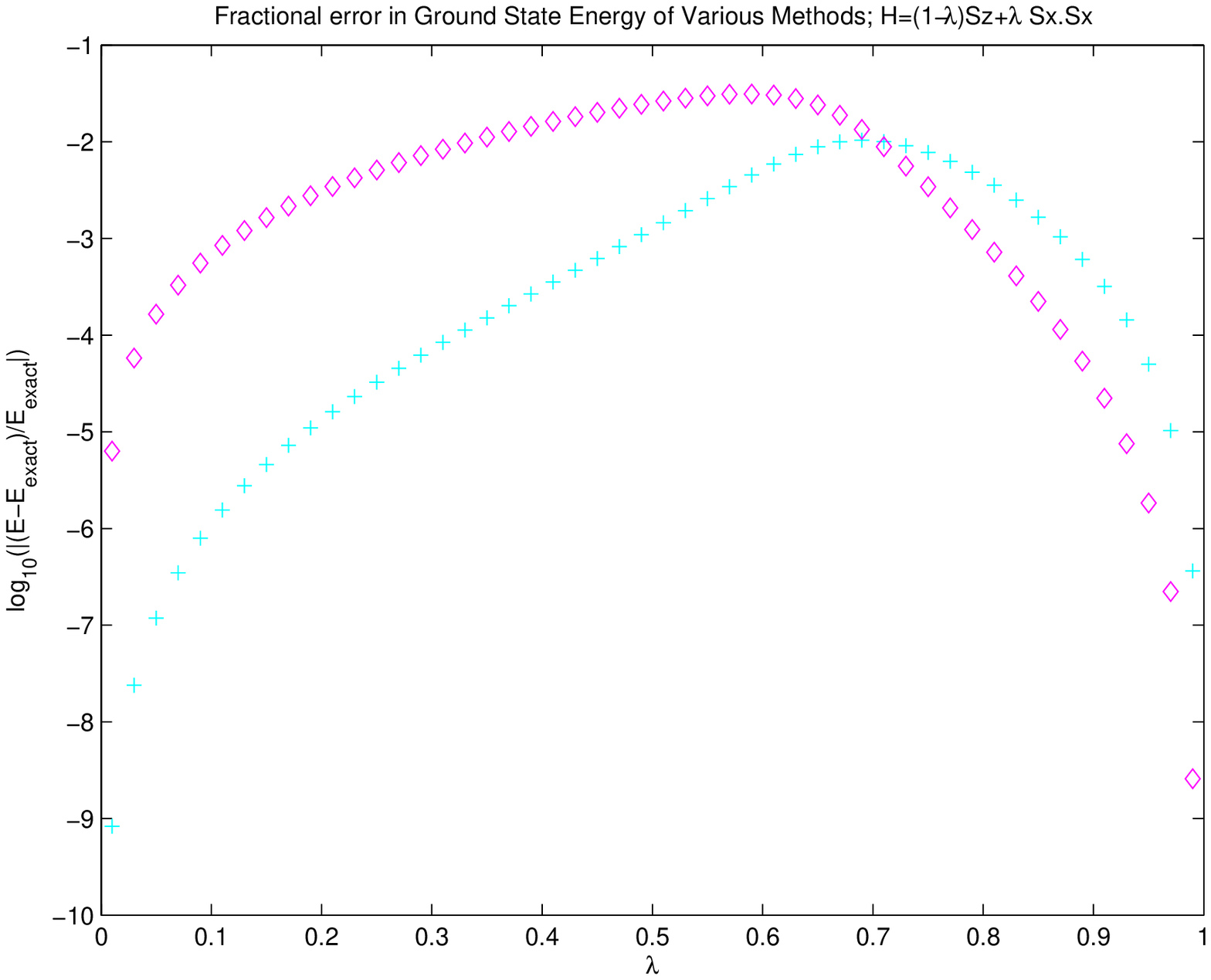}\\
  \includegraphics[width=3.4in]{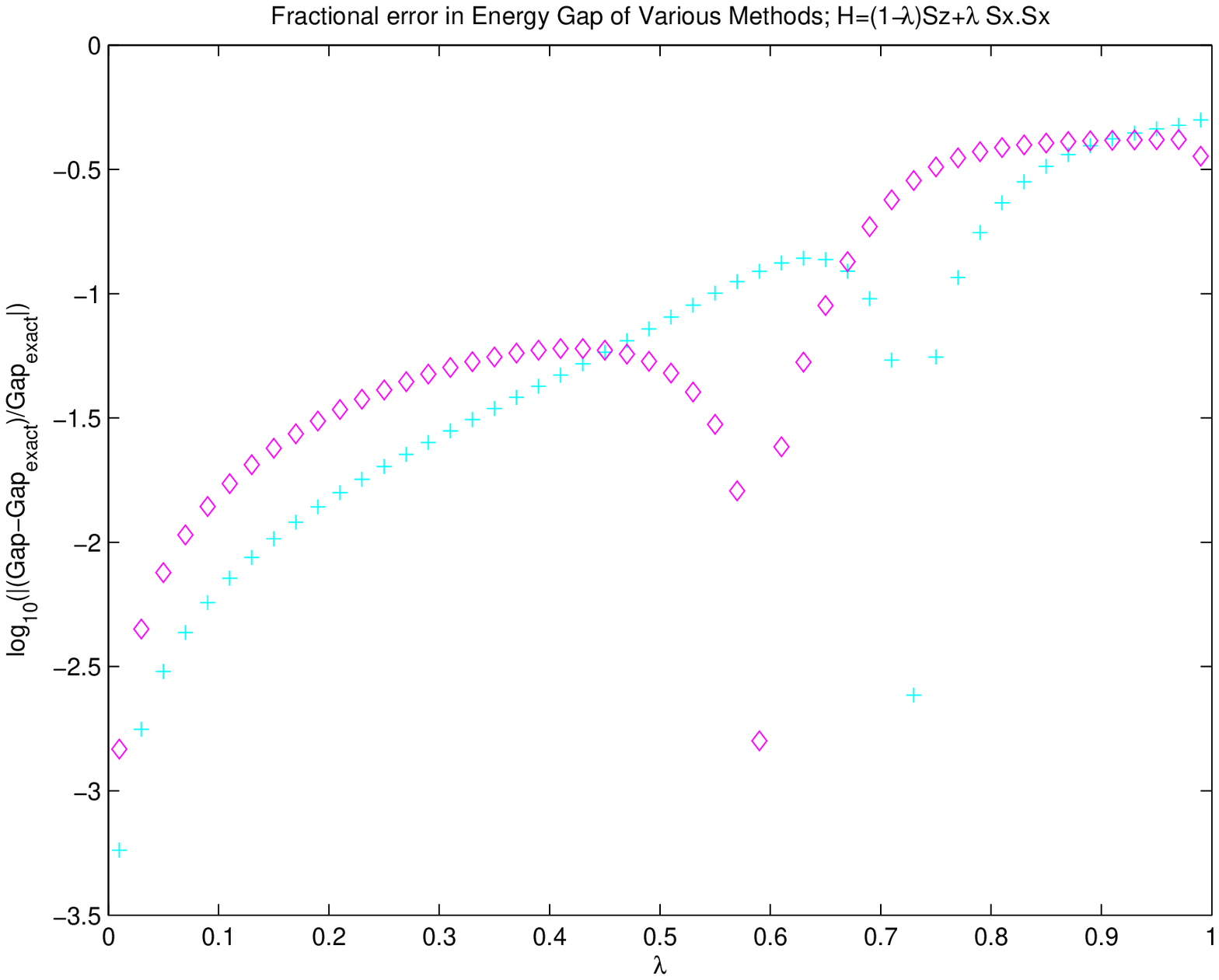}\\
  \includegraphics[width=3.2in]{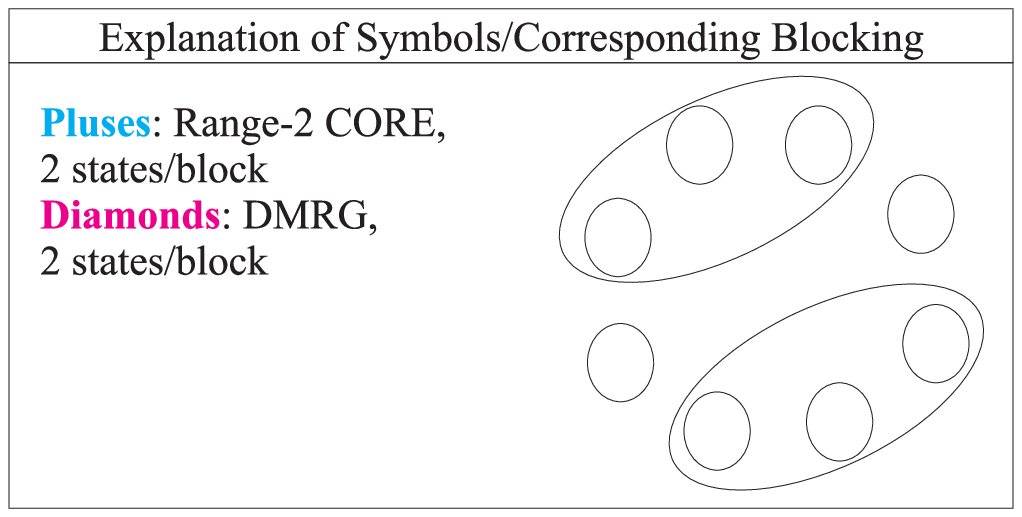}\\
  \caption{Comparison of CORE to DMRG on an eight-site periodic Ising system. The blocking mimics the intermediate configuration of a DMRG sweep with periodic
boundary
  condition.}
\label{isingb}
\end{figure}

Fig.\ref{ising} shows that CORE compares favorably against HDMRG, so
let us next compare CORE to DMRG with a blocking scheme that is
natural to the latter. The chosen system is an eight-site periodic
Ising chain with two blocks and two free sites, shown in
Fig.\ref{isingb}. The two three-site blocks mimics the system/
environment in the middle of a DMRG sweep, and the two free sites
are positioned according to the prescription in Ref.\cite{periodic}.
Two states are kept in each block. As in the HDMRG case, we simulate
the iteration (again we refer the readers to
Refs.\cite{densitymatrix, periodic} for details) by using the exact
ground state as our target state, so our result represents the
convergence limit. If we had actually carried out the iterative
sweeps, we would need to diagonalize at most $2^4$ states at a time
(two states in each free sites, two states in the system block and
two states in the environment block), so for fairness, we calculate
CORE only up to range-2 because it also requires diagonalization of
$2^4$ states (one block plus one free site exactly). Here we simply
use the first version of CORE which truncates to the lowest energy
in each block. Plots in Fig.\ref{isingb} show that both DMRG and
CORE achieve very high accuracies, but the two methods excel at
different regimes of the Ising system. This indicates that CORE's
performance, at least in simple situations, is comparable to DMRG
without any need for iteration.

\subsection{b. CORE as a disentangling algorithm}

The presence of entanglement is often believed to be what makes
quantum systems difficult to simulate. By "entanglement" we mean a
measure that quantifies how much a quantum state deviates from a
tensor product of states on subsystems. In the case of a pure state,
for example, we can measure how much a finite block is entangled to
the rest of the system using its von Neumann entropy, \be
S(\rho)=-\Tr(\rho \ln \rho) \label{vne} \ee where $\rho$ is the
reduced density matrix of the block. Efficient preservation of
entanglement accounts for much of DMRG's success\cite{QIDMRG}.

Without explicit consideration of entanglement, how does CORE
achieve a performance comparable to DMRG? In this subsection we will
try to understand this through theoretical comparison with an
entanglement-based method.

Vidal recently proposed a method called Entanglement Renormalization
(ER) \cite{DU}, which we can think of as an improvement on HDMRG.
Recall that, by HDMRG, we refer to the method that uses reduced
density matrix truncation as DMRG does but blocks hierarchically as
CORE and naive renormalization do. Vidal observed that before we
truncate states in a block according to the reduced density matrix
of a target state, it is possible to apply a unitary transformation
on the boundary sites of two adjacent blocks and reduce the
entanglement of the target state. In other words, the transformation
makes states which are truncated away have less weight in the
reduced density matrix of the block. Thus the same number of
retained states can preserve more of the information of the original
system. In exchange, we pay the price that operators acting on one
block act on neighboring blocks after the disentangling unitary
transformation. The renormalized Hamiltonian can be written as:
\begin{eqnarray}
H^{ER}_{ren}=W^\dag UHU^\dag W, \nonumber \\
U=U_{1,2}\otimes U_{2,3} \otimes...\label{DUeq}
\end{eqnarray}
where the $\{U_{i,i+1}\}$'s are disentangling unitary
transformations acting on the edges of neighboring blocks $i, i+1$
and $W$ is an isometric transformation ($W^\dag W=I$) lifting from
the coarse-grained subspace to the full Hilbert space. For example,
if $\{\ket{a_i}\}$ form a basis of the subspace and $\{\ket{b_i}\}$
form a basis of the full space, we can write
$W_{ij}=\iprod{b_i}{a_j}$, i.e. a matrix whose columns are the basis
vectors of the subspace. The orthogonal projection $P$ is related to
$W$ by $P=\sum_k \ket{a_k}\bra{a_k}=WW^\dag$.

Like CORE, Entanglement Renormalization stores some information in
operators in addition to the retained states. The two methods turn
out to be have a similar starting point. In Section 2 we have shown
the form of the renormalized Hamiltonian in the limit of
$t\to\infty$. Now, since both $\{\ket{w_j}\}$ and
$\{\ket{v_{f(j)}}\}$ (defined in Lemma~\ref{lemmaofR}) are
orthonormal, we can think of the mapping which identifies each
$\{\ket{w_j}\}$ with its corresponding $\{\ket{v_{f(j)}}\}$ as the
restriction of a unitary transformation which acts on the full
Hilbert space.  Of course, given our construction we only have the
restriction of this transformation to the subspace spanned by the
retained states, the extension of the transformation to the full
Hilbert space remains undefined and is presumably not unique. The
fact that this unitary transformation is not uniquely specified is
equivalent to the fact that $W$ and $W^\dag$, which appear in the ER
formulas, are isometries and not unitary transformations.

To put the correspondence between CORE and ER in a formal setting we
make the following claim:
\begin{claim}
For all unitary operators $U$ such that $U\ket{v_{f(j)}}=\ket{w_j}$,
\be H_{ren}=PUHU^\dag P. \label{PU} \ee
\end{claim}
\begin{proof}
\begin{eqnarray}
PU&=&\sum_{i}\ket{w_i}\bra{w_i}U \nonumber \\
&=&\sum_{i,l}\ket{w_i}\bra{w_i}U\ket{v_l}\bra{v_l} \nonumber \\
&=&\sum_{i,l}\ket{w_i}\delta_{f(i)l}\bra{v_l} \nonumber \\
&=&\sum_{i}\ket{w_i}\bra{v_{f(i)}} \nonumber \\
\end{eqnarray}
where the third line follows from the observation that $U\ket{v_l}$
must be orthogonal to $\cal H'$ if $\forall i,~l\neq f(i)$. Hence we
can write \bea PUHU^\dag
P&=&\sum_{i,j}\ket{w_i}\bra{v_{f(i)}}H\ket{v_{f(j)}}\bra{w_j}
\nonumber \\
&=&\sum_{i} \ket{w_i} E_{f(i)} \bra{w_i}\eea which is exactly
Eq.\ref{diagform}.
\end{proof}

As we noted earlier, we can write $P=WW^\dag$. In practice, we want
to write $H_{ren}$ in a basis of the subspace $\cal H'$, so what we
really calculate is $H_{ren}=W^\dag UHU^\dag W$, just as in
Eq.\ref{DUeq}. Thus, we see that both CORE and Entanglement
Renormalization use a renormalized Hamiltonian of the form
$PUHU^\dag P$. The distinction between CORE and ER is that
\emph{Entanglement Renormalization approximately disentangles a
system, while CORE approximates a disentangled system}. We say that
Entanglement Renormalization approximately disentangles the system
because there is no guarantee that we can find disentangling
unitaries that reduce the rank of the reduced density matrix to less
than or equal to the dimension of retained states.  It is usually
necessary to truncate some states to keep the degrees of freedom
manageable, and information is lost when we truncate the states.
CORE approximates a disentangled system in the sense that we first
write down the form of a completely disentangled system
(Eq.\ref{PU}). Its cluster expansion truncated to diameter-$k$ then
approximates this system by ensuring that the new Hamiltonian
restricted to any sublattice with diameter less than $k$ is
completely disentangled. Here information is lost when we truncate
the operators with diameter larger than $k$.

In this picture, the CORE prescription for choosing $R$ and
$E_{f(i)}$ is a particular choice of disentangler.  One might
conceive of a disentangler $PU$ that does not always require
overlaps between $\ket{w_i}$ and $\ket{v_{f(i)}}$  - a trivial
example would be to set $H_{ren}=\sum_{i} \ket{a_i} E_{i} \bra{a_i}$
where $\{\ket{a_i}\}$ is the basis of $\cal H'$ that we start with.
The problem with such an arbitrary choice is that the exact
renormalized Hamiltonian could be so non-local that it is difficult
to approximate it by a cluster expansion.  Our intuition is that by
keeping $\ket{w_i}$ somewhat "close" to $\ket{v_{f(i)}}$, cluster
expansion can do well. There may be, however, room for improvement
once we have a better understanding of the cluster expansion.

\subsection{c. The use of entropy in CORE}

\begin{figure}
  \includegraphics[width=3.4in]{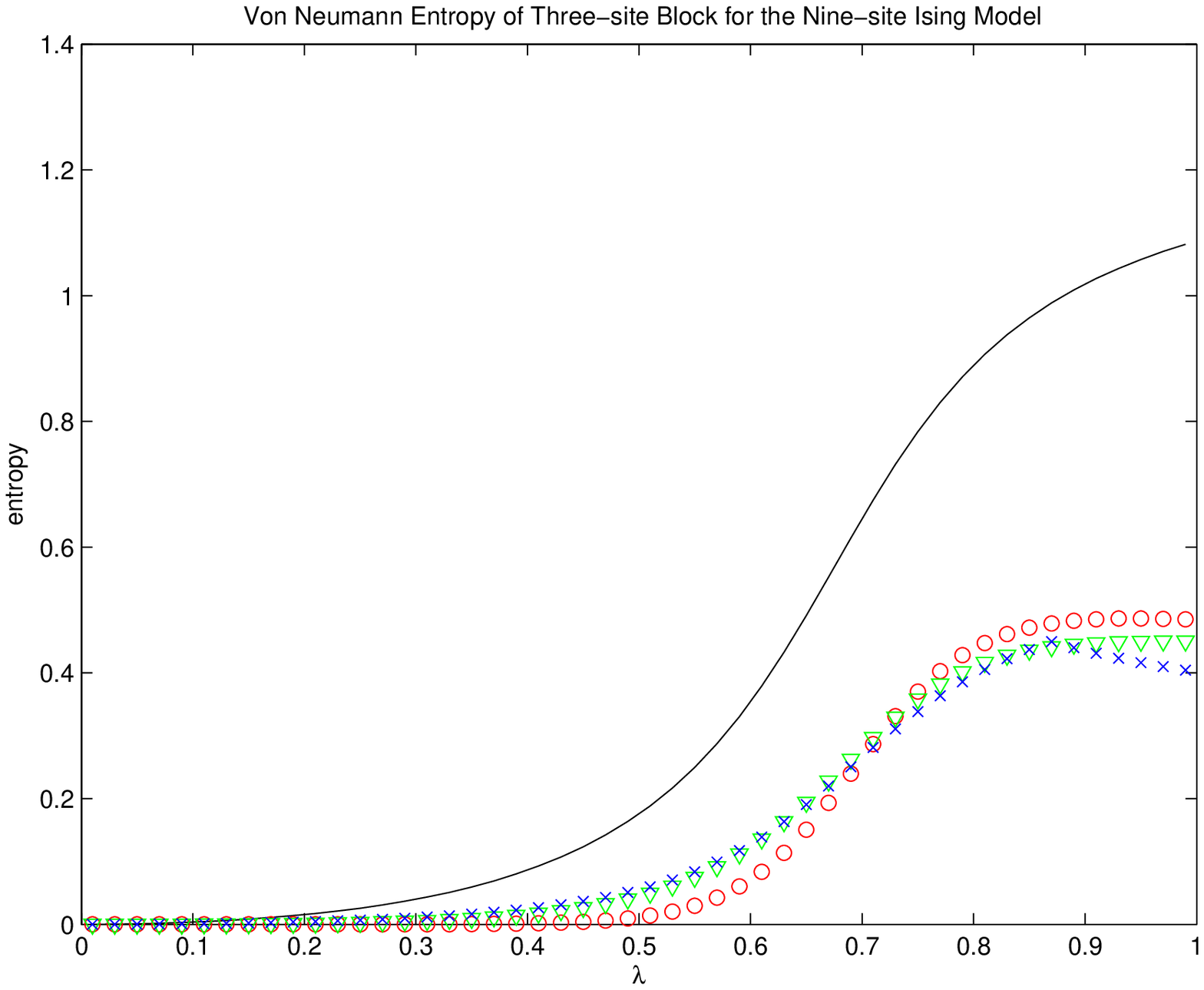}\\
  \includegraphics[width=3.4in]{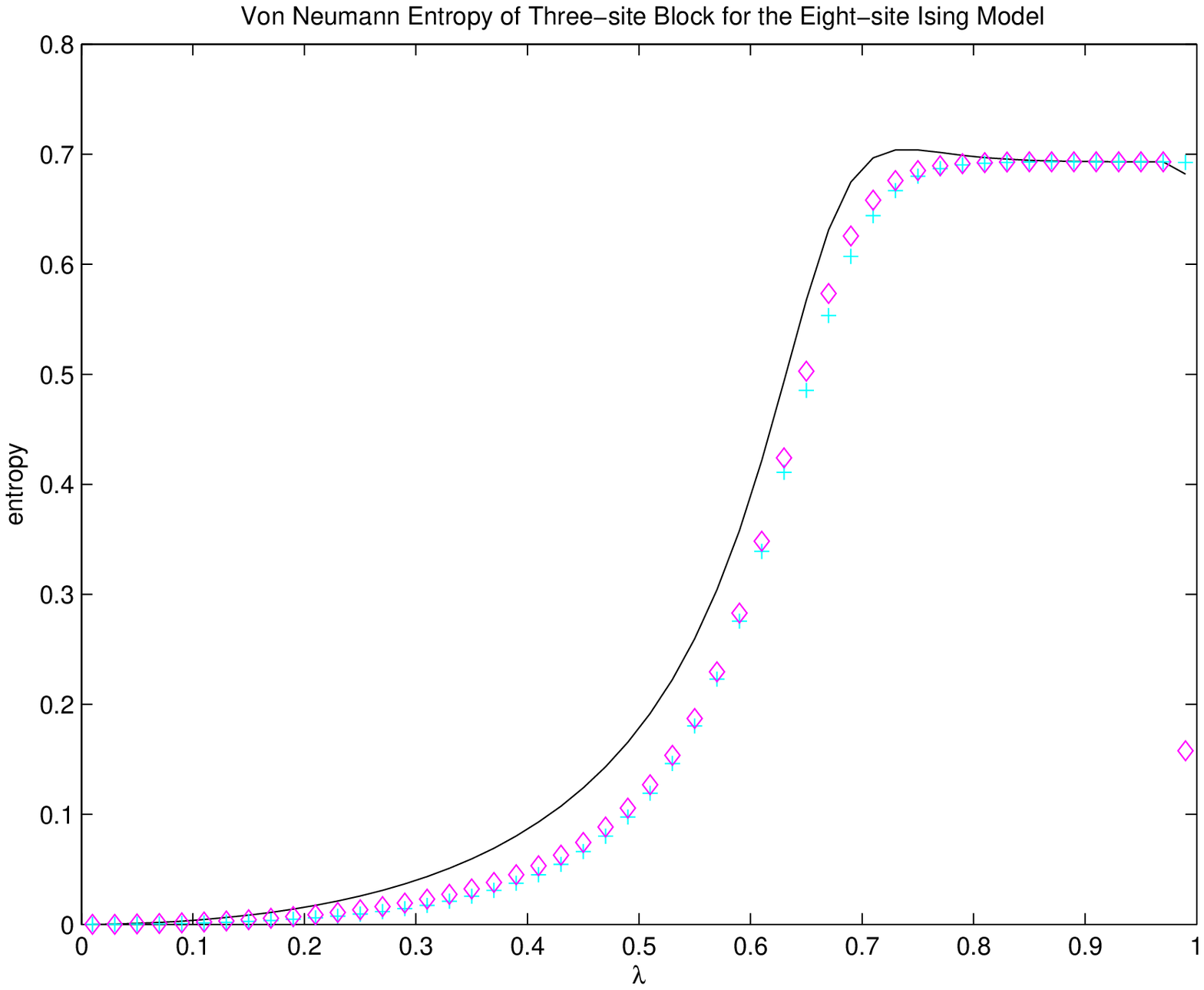}\\
  \includegraphics[width=3.4in]{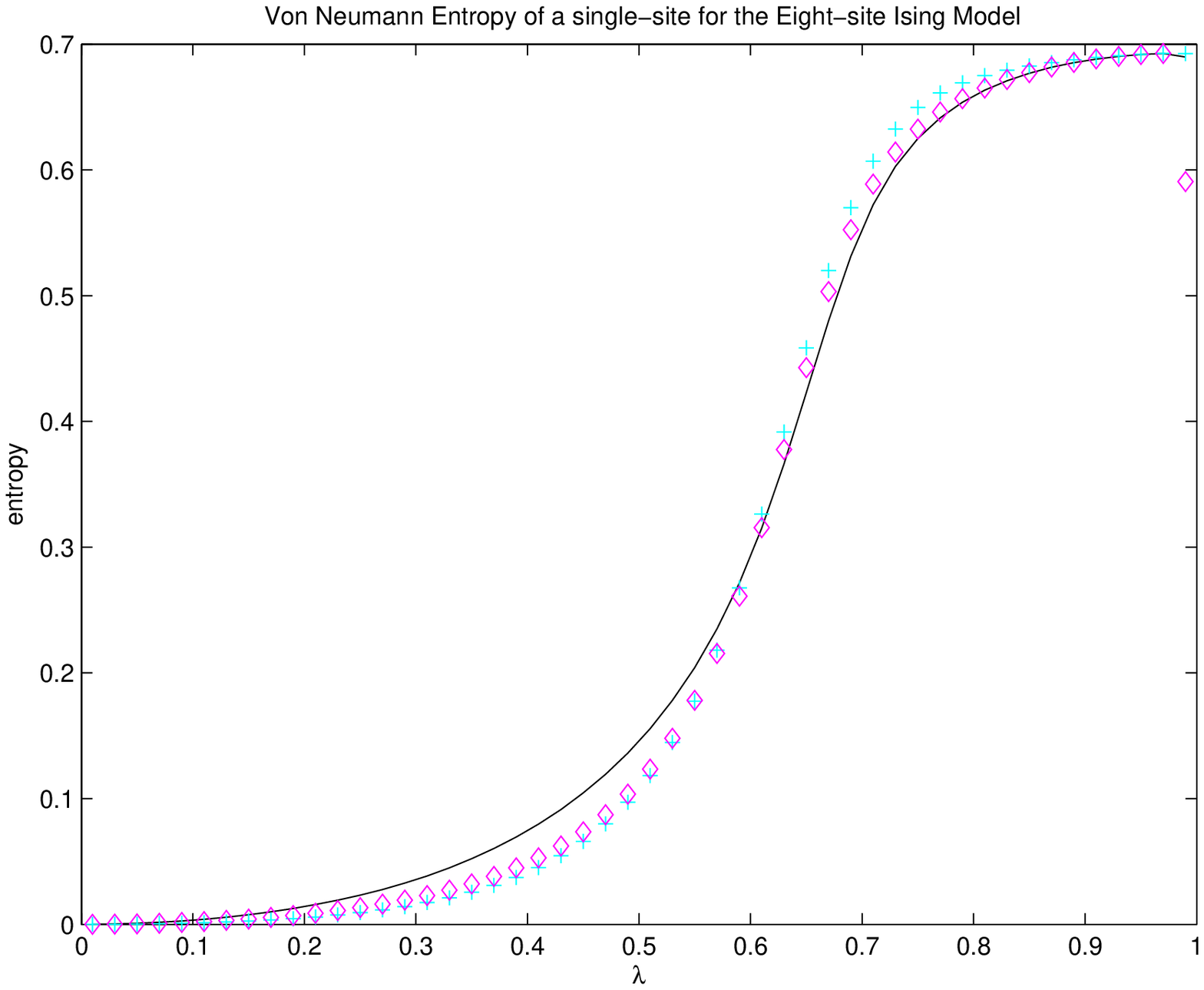}\\
  \caption{Here we show the block entropy as a function of coupling for the
  nine-site and eight-site Ising models. The solid curve indicates the
  exact entropy and other symbols represent methods explained in previous
  plots. The first two plots are for three-site blocks and the last
  for a single-site. We see that CORE reproduces the shape in a manner similar to DMRG.}
\label{isingent}
\end{figure}

The von Neumann entropy (Eq.\ref{vne}) of finite blocks in a lattice
is known to exhibit scaling behaviors with block size that depend on
the dimension and phase of the system \cite{Latorre}. Because DMRG
truncates according to the reduced density matrix, entropy can be
approximately preserved, so apart from using entropy as a measure of
how much information is lost, it is also possible to use the
approximate entropy measures from DMRG to detect phase transitions
\cite{LegezaPT}.

Does entropy have a similar use in CORE? Note that CORE is not a
variational method that works with a subspace within the original
space. It tries to preserve eigenvalues but not the eigenvectors and
this allows a disentangling effect. Therefore there is no \em a
priori \em reason to expect CORE to preserve any particular amount of
entropy.

Nonetheless, when we evaluate the three-site block entropy for the
two toy models in Section 4(a), we were impressed by how much
information CORE captures. Fig.\ref{isingent} shows the entropy of
some blocks within the eight-site and nine-site models (the symbols
are explained Fig.\ref{ising} and Fig.\ref{isingb}). The first two
plots are for a three-site block, in which the eight states are
truncated to two after the renormalization. This means that the
upper bound of the block entropy changes from $\ln 8$ to $\ln 2$. As
it turns out the exact block entropy in the eight-site case is far
from saturating the bound, allowing DMRG and CORE to approximate it
closely. The third plot in Fig.\ref{isingent} is for the single-site
entropy which is available only for the eight-site configuration
(The sum of single-site entropies can be used to define the total
information encoded in the wavefunction\cite{LegezaEnt}). Here there
is no truncation in the site itself, so the entropy upper bound is
the same before and after the renormalization.

It may seem confusing how CORE can have a disentangling effect and
yet keeps a block entropy that's comparable to DMRG. By
"disentanglement" we mean a reduction of weight on the eigenvalues
of the density matrix outside of the retained space, which does not
say anything about the distribution of eigenvalues inside the
retained space. It is the latter that determines the block entropy
after truncation.

What is remarkable about Fig.\ref{isingent}, however, is not the
amount of entropy CORE keeps, but the fact that it varies smoothly
with the coupling $\lambda$ and has a shape similar to the exact
entropy. This raises the possibility that entropy in CORE can also
detect quantum phase transitions. Regrettably, unlike Entanglement
Renormalization we cannot perform disentangling unitaries without
truncation to check entropy scaling in critical and noncritical
systems\cite{DU}, which means our result can be highly dependent on
our choice of retained states and cluster expansion. In absence of
more data, we are hesitant to draw conclusions at this point, but
this can be an interesting area to explore.

\subsection{d. The choice of retained states}

In Fig.\ref{ising}, we have shown two versions of CORE with
different choices of retained states, one of which is inspired by
DMRG. Although we have not done so, for the study of entropy we just
discussed one can even try other variants of DMRG that choose
retained states by maximizing entropy \cite{QIDMRG} or minimizing
the Holevo $\chi$ \cite{LegezaEnt}. How choices of retained states
affects the performance of CORE have never been investigated in the
past so we would like to briefly discuss it here.

In the single-block calculation it is clear that whatever states we
retain, as long as they have an overlap with the lowest lying states
of the block, the renormalized range-$1$ (i.e. single site)
Hamiltonian will be unchanged. It is only when we compute the
connected range-$2$,range-$3$, etc., operators that the difference
in choice of retained states show up.  In general, the effect of
changing the choice of single block retained states can either
increase or decrease the size of the longer range connected
interactions.

Past works on CORE have exclusively retained states with the
smallest energy in the single block Hamiltonian. Let us call this
the first version of CORE. Inspired by DMRG, we have tried a second
version that uses states with the largest eigenvalues in the reduced
density matrix of some target state. When we choose the target state
to be the exact ground state of two adjacent blocks, we often obtain
a slightly better ground state energy at the expense of a less
accurate gap between the ground state and the first excited state,
as can be seen in Fig.\ref{ising}. When we choose the target state
to be the exact ground state of a larger block that contains our
block in the center, however, the retained states often turn out to
be the same as those in the first version, i.e., the states with the
smallest single block energy. This latter observation is in
agreement with the results of Capponi et al \cite{Capponi}, who
found that the retained states in the first version have very high
weights in the reduced density matrix of a target state of a larger
block.

Capponi et al considered this question in a different context and
proposed that one should use the reduced density matrix to check if
the truncation in the first version of CORE is justified. We think
that such a diagnostic tool has to be used very cautiously. To see
why we say this, consider first the case when the target state is
calculated on two adjacent blocks. Here the diagnostic tool
essentially measures the overlap between $\ket{v_{f(1)}}$ and
$\ket{w_1}$ (on the Hilbert space of two blocks) in Eq.\ref{PU}. A
large overlap, however, does not guarantee that $U$ is close to the
identity, let alone a better convergence of the cluster expansion.
An alternative is to use the mixed state corresponding to
combination of all $\{\ket{v_{f(j)}}\}$ as the target state, which
should more meaningfully measures how much $U$ differs from the
identity. Yet it still does not in general reliably indicate how
well the cluster expansion converges. All we can be certain of is
that in the limit $U\to I$, the cluster expansion tends to be exact
at nearest neighbor range; this does not tell us much when $U$ is
significantly different from $I$. In particular, if we consider the
case when the target state is not the ground state on two blocks,
but on a larger block containing the block we want to truncate, the
situation is even less clear because in that case CORE does not
explicitly rotate this ground state to the retained states.

The reduced density matrix tells us a lot in DMRG, but as we have
mentioned in the previous subsections, CORE is performing some
disentangling action and thus does not preserve the entanglement
structure of the states as DMRG does. Simply preserving the most
entanglement in the original Hilbert space does not guarantee the
most accuracy in CORE. Fortunately, it turns out that in our Ising
example, when we choose our retained states to maximize
$\iprod{v_{f(1)}}{w_1}$ on two blocks, we do get a better overall
ground state (and sacrifice the accuracies of the excited states).
This seems to indicate that the reduced density matrix can be
helpful - as long as we remember that even as
$\iprod{v_{f(1)}}{w_1}$ goes to 1, there is no substitute for the
longer-range terms in the cluster expansion.

\section{6. Discussion and Further Directions}

This paper has presented a number of numerical results on various
models. While they have shown an usefulness consistent with the
considerable body of literature on CORE/RSRG-EI, the picture is
still far from clear. The fact that we get an improvement in energy
from longe-range operators in 2-D shows non-trivially that there is
a genuine small parameter hidden in the cluster expansion, which we
believe is associated with the \em diameter \em and not the number
of blocks in the configuration. Yet we still have little knowledge
how this convergence manifests itself under different blocking
schemes. When we tested the five-site blocking shown in Section
3(c), we expected the growing spin picture to be valid on prior
physical grounds, so we were surprised by how much the long-range
terms could affect the result. There seems to be a need for a way to
at least estimate the effect of long-range operators. As far as we
know, only a very few applications of CORE/RSRG-EI have considered
long-range operators up to "diameter-$\sqrt{2}$" \cite{calzado}.
This is presumably due to limited computational power. Is it
possible to use other approximate methods of diagonalization, such
as perturbation theory or DMRG, to estimate these long-range
operators? This hybrid approach is certainly a direction we would
like to pursue in the future.

The second part of the paper, where we compare CORE to
entanglement-based approaches, also raises a number of questions. In
Section 4(b), we showed that CORE is in fact theoretically similar
to Entanglement Renormalization. Naturally, it would be interesting
to see how ER performs numerically in comparison to CORE, and in
particular, whether one can reproduce the three pictures of the
antiferromagnet with ER. Section 4(c) raises the possibility that
apart from the energy spectrum, we may be able to use the entropy to
study phase transitions with CORE. This, if true, would be
remarkable given that CORE was not designed to preserve entropy. In
Section 4(d) we mentioned the use of reduced density matrix as an
alternative method of deciding what states to keep in each block.
There is a special circumstance when it can be important. This has
to do with a situation which comes up as soon as one studies
frustrated HAF's where, with increasing frustration, single block
levels tend to cross.  Since one expects that the most interesting
physics of these models is associated with these regions one has to
know what states to choose when degeneracies occur.  The most
obvious solution is to keep all states which can cross as a function
of the truncation, however this will increase the complexity of the
numerical computations. In that case it may be advantageous to
retain the states determined by a DMRG calculation which have the
highest weight and correct spin.

Finally, it would be remiss of us not to point out, without giving
details, the possibility of adapting the principles of CORE to other
applications. We noted in Section 4 that specific truncation and
blocking methods are details of the projection operator; the
underlying principle is a general one about simplifying states at
the expense of generating non-local operators. The principles of
DMRG have been generalized to simulate real time
evolution\cite{TEBD} - Can the principles of CORE can be applied to
time evolution as well? There are good motivations for thinking
about this. One of the most efficient simulation method for a
specific class of unitary evolutions is the stabilizer formalism
\cite{stabilizer}, where we do not keep explicit information about
the states and instead keep track of them using a set of operators.
Since CORE allows one to calculate expectation values at the expense
of state information, we could ask if cluster expansion can be
similarly efficient for certain types of unitary evolution. Apart
from a direct simulation of unitary evolution, we may also consider
turning a unitary evolution problem into a ground state problem
using results in quantum complexity theory\cite{kitaev} (Linear,
instead of hierarchical, blocking would have to used in that case).
All such possibilities would be interesting to explore.

\section{Acknowledgement}
We would like to thank Samuel Moukouri for valuable exchanges.

\vskip 15pt
\appendix
\noindent    {\bf Appendix: QR-Decomposition and Proof of Lemma
\ref{lemmaofR}} \vskip 5pt

While constructive proof of Lemma \ref{lemmaofR} has been given in
the past\cite{COREpaper}, for practical implementation, we have
found recursive QR-Decomposition to be a fast and convenient way of
calculating the rotation $R$ (particularly when packaged libraries
such as NAG or LAPACK are available). Given any $M \times N$ matrix
$A$, $M\leq N$, the QR-Decomposition is defined by ${\cal QR=A, Q^T
Q=}I$ where $\cal Q$ is an $M \times M$ matrix and $\cal R$ is an
upper-triangular $M \times N$ matrix. For our application, $A$ is
simply the overlap matrix $O_{il}$, and the matrix $\cal Q$ plays
the role of $R^\dag$ (Thus the $\cal R$ of $\cal QR$ is not the $R$
of Lemma \ref{lemmaofR}). Note that the upper triangular matrix does
not necessarily satisfy the contraction condition
(Eq.\ref{contract}); it merely guarantees zeros entries below the
diagonal $\{{\cal R}(i,i)\mid i=1..M\}$, a particularly weak
condition if $M\ll N$. Our solution is to start from the upper-left
corner, move down the row and column after every QR-Decomposition
until we find another non-zero entry that has non-zero entries below
it (this means more than one retained states contract to the
eigenstate), at which we perform another QR-Decomposition on the
submatrix with that entry as the upper-left corner. We repeat this
recursively until the submatrix size is zero.


\end{document}